\documentclass[conference,onecolumn]{IEEEtran}
\usepackage{algorithmicx}
\usepackage{algpseudocode}
\usepackage{algorithm}
\usepackage{array}
\usepackage[caption=false,font=normalsize,labelfont=sf,textfont=sf]{subfig}
\usepackage{textcomp}
\usepackage{url}
\usepackage{verbatim}
\usepackage{graphicx}
\usepackage{cite}
\usepackage{amsmath,amssymb,amsfonts}
\usepackage{balance}
\usepackage[left=0.68in,right=0.673in,top=0.7in,bottom=1.013in]{geometry}

\usepackage{subcaption}
\usepackage{xcolor}
\usepackage{caption}
\usepackage[normalem]{ulem}
\usepackage{makecell}
\usepackage{tikz}
\usetikzlibrary{arrows.meta,positioning,fit}

\begin{document}

\title{Tackling Interference in HAPS Networks via Angular-Aware Clustering and RSMA}

\author{\IEEEauthorblockN{Afsoon Alidadi Shamsabadi\IEEEauthorrefmark{1},
Animesh Yadav\IEEEauthorrefmark{2}, and Halim Yanikomeroglu\IEEEauthorrefmark{1}}
\IEEEauthorblockA{\IEEEauthorrefmark{1}Carleton University, Ottawa, ON, Canada, and \IEEEauthorrefmark{2}Ohio University, Athens, OH, USA\\
Email: afsoonalidadishamsa@sce.carleton.ca, yadava@ohio.edu, halim@sce.carleton.ca}}

\maketitle

\begin{abstract}
High Altitude Platform Stations (HAPS) have emerged as a promising enabler for next-generation wireless networks, offering ubiquitous connectivity to ground users. Operating either in standalone mode or in integration with terrestrial networks, HAPS can significantly enhance both coverage and capacity due to their strategic placement in the stratosphere. However, interference management in HAPS-empowered networks requires special attention due to the unique propagation characteristics of HAPS links. In particular, the strong line-of-sight (LoS) conditions between HAPS and ground users result in limited channel variability, thereby intensifying inter-user interference. In this work, we consider a single HAPS serving multiple ground users through multiple beams over a limited number of orthogonal resource blocks (RBs). To address the resulting interference, we propose a novel angular-aware user clustering and interference-aware RB allocation framework that strategically clusters users, designs beams to serve each cluster, and allocates RBs to users across clusters. To further mitigate intra-RB interference, a rate-splitting multiple access (RSMA) scheme is incorporated. Simulation results demonstrate that the proposed clustering and RSMA-based approach significantly outperforms baseline schemes in terms of achievable per-user spectral efficiency.
\end{abstract}

\section{Introduction}\label{Introduction}
Next-generation (xG) wireless networks will face critical challenges in terms of coverage and capacity. Under the \textit{ubiquitous connectivity} paradigm for sixth-generation (6G) and beyond networks, standalone terrestrial networks will not be able to meet these stringent requirements. Therefore, xG networks are expected to evolve toward integrated terrestrial and non-terrestrial networks (NTN), where multiple network tiers, including space, aerial, and terrestrial, cooperate to serve both rural and urban users on the ground~\cite{6GRoad}. Among aerial platforms, high altitude platform stations (HAPS), which are quasi-stationary access nodes located in the stratosphere, are promising solutions for complementing terrestrial networks. When integrated with terrestrial networks in overlapping coverage regions, forming vertical heterogeneous networks (vHetNets), HAPS can improve network performance by serving users who (i) experience poor quality-of-service (QoS) from terrestrial networks or (ii) cannot be served due to congestion at terrestrial macro base stations (MBSs). The HAPS tier can operate in either the same spectrum as, or a different spectrum from, the terrestrial tier. When sharing the same spectrum (forming a harmonized-spectrum vHetNet), the integrated network faces severe inter- and intra-tier interference, making efficient interference management strategies essential~\cite{IM_Magazine}. 

To this end, the authors in~\cite{Ishikawa} proposed a resource-allocation-aided null-forming scheme for HAPS-terrestrial spectrum sharing, where nulls are selectively disabled over certain time-frequency resources and terrestrial MBSs adapt user scheduling accordingly, thereby mitigating coverage holes and improving signal-to-interference-plus-noise ratio (SINR) performance. Moreover, the authors in~\cite{Kirik} propose a three-phase interference mitigation framework for HAPS-assisted 6G networks, leveraging subspace-based blind channel estimation, optimized sparse codebook design, and pilot-assisted transmission to suppress co-channel interference. Additionally, in~\cite{Our-WCL}, we developed a joint subcarrier allocation and power control scheme to manage interference in an integrated HAPS--terrestrial network. Accordingly, in~\cite{our-ICC} and~\cite{our-CL}, we developed centralized user association and beamforming algorithms to tackle interference in a harmonized-spectrum vHetNet. These works were further extended in~\cite{TwoLevel}, where we developed a low-complexity distributed beamforming weight design algorithm for a cell-free harmonized-spectrum vHetNet.

Due to their unique geometry and propagation environment, HAPS networks exhibit distinct interference characteristics. In particular, HAPS covers a wide geographical area and serves a large number of user equipment (UEs). Due to its high altitude, HAPS–UE links are typically dominated by strong line-of-sight (LoS) conditions. Moreover, although UEs may be geographically well separated from a terrestrial-network perspective, they appear closely spaced from the HAPS viewpoint. This results in highly similar channel conditions among UEs, making interference management a critical challenge. To this end, various strategies such as resource allocation, antenna architectures, beamforming, etc., can be employed to mitigate the interference in HAPS networks. Furthermore, advanced multiple access schemes such as rate-splitting multiple access (RSMA) have recently emerged as powerful tools for interference management, as they enable flexible decoding of interference and provide a bridge between treating interference as noise and fully decoding it~\cite{RSMA-Popovski}. 
As in \cite{Mehmet}, the authors integrate RSMA with multi-numerology orthogonal frequency-division multiplexing (OFDM) to jointly address multi-user and inter-carrier interference, formulating a weighted minimum mean-square error-based optimization for power and subcarrier allocation that improves sum-rate and fairness compared to OFDM and non-orthogonal multiple access (NOMA). Moreover, the authors in \cite{Zhang} propose a distributed RSMA-based approach with deep learning to manage interference in space–air–ground integrated networks, achieving improved sum-rate performance with significantly reduced computational complexity, and the authors in \cite{Huang} propose a robust RSMA-based secure precoding scheme for HAPS-assisted satellite networks, jointly optimizing satellite and relay transmissions under imperfect channel state information to enhance secrecy rates compared to conventional multiple access schemes.

In this work, we focus on interference management in HAPS networks through UE clustering, RB allocation, and RSMA. In contrast to prior works where HAPS serves UEs through UE-specific beams, we assume that HAPS forms beams toward geographic areas of clusters, with multiple UEs served by the same beam. This assumption is more practical compared to the UE-specific beams, given the HAPS altitude and the relative proximity of UEs from the HAPS viewpoint. Accordingly, we develop an algorithm to cluster UEs based on their angular positions. We then propose an interference-aware scheme to allocate RBs to UEs so that RBs are allocated orthogonally within each cluster to mitigate intra-cluster interference; however, inter-cluster interference between UEs sharing the same RB remains a challenge. To address this, we employ RSMA on each RB to manage interference among UEs sharing that RB. Together, the proposed clustering, RB allocation, and RSMA transmission strategies effectively manage interference in HAPS networks and can enhance the performance of the UEs in terms of spectral efficiency (SE).

The rest of the paper is organized as follows. Section~\ref{Sec:System Model} presents the system model. Section~\ref{Sec:Clustering and RB Allocation} introduces the angular-aware clustering and interference-aware RB allocation. Section~\ref{Sec:RSMA} describes the proposed RSMA-based interference management scheme. Section~\ref{Sec:MMFProblem} formulates the max-min fairness (MMF) optimization problem to design the common and private parts' powers and proposes an iterative algorithm to solve the problem. Section~\ref{Sec:SimResults} provides numerical results, and Section~\ref{Sec:Conclusion} concludes the paper.

\section{System Model}\label{Sec:System Model}
\begin{figure}[t]
    \centering
    \captionsetup{justification=centering}
    \includegraphics[width=0.7\linewidth]{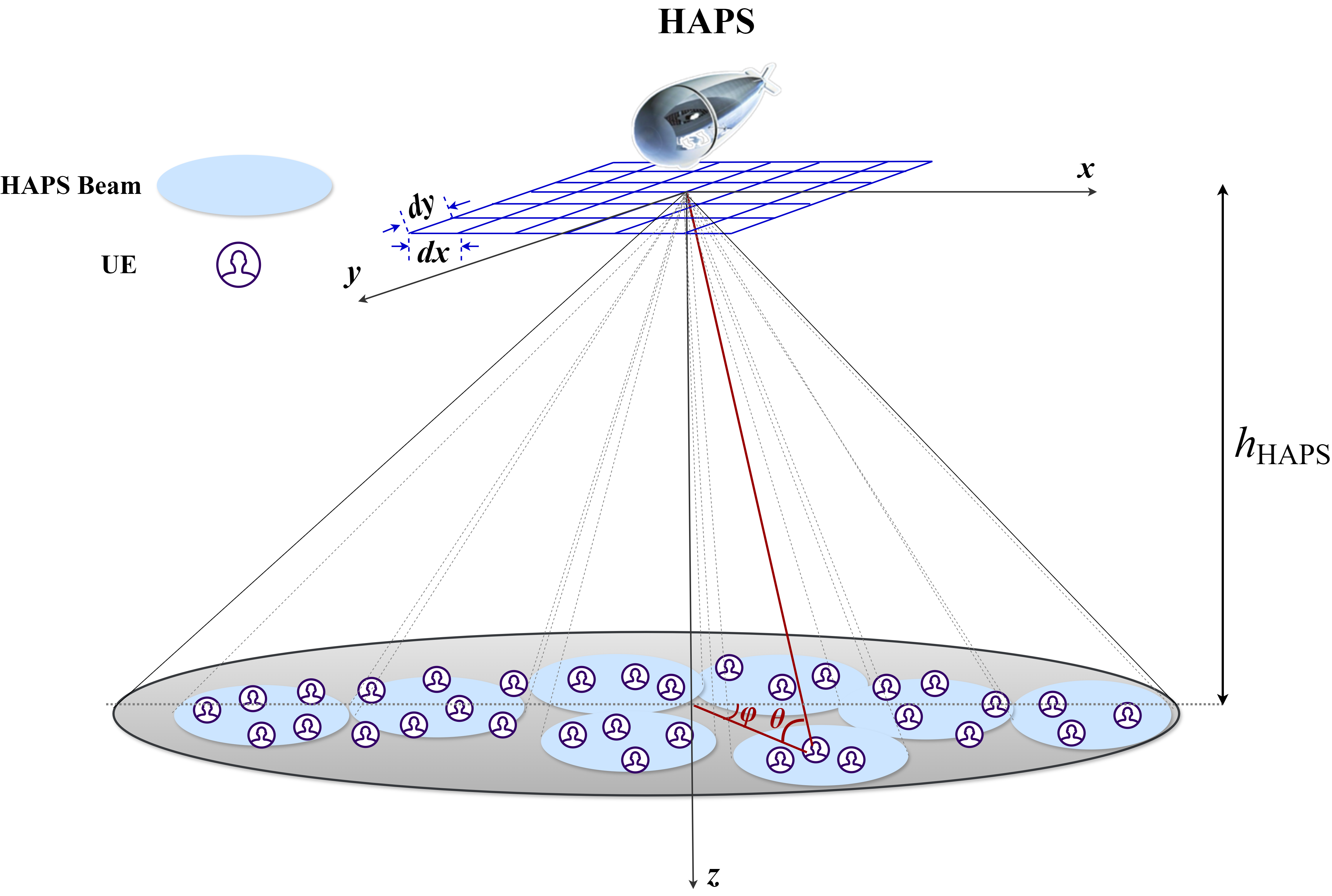}
    \caption{\small System model.}
    \label{systemmodel}
\end{figure}

We consider a single HAPS providing coverage over a geographical area of size $A \times A \text{ km}^2$, serving $U$ single-antenna UEs in downlink channels over $R$ orthogonal RBs, as depicted in Fig.~\ref{systemmodel}. We assume that the HAPS and terrestrial networks operate on different frequency bands, thereby eliminating interference between the tiers. However, there will be interference among the UEs served by HAPS, leading to intra-tier interference. The sets of UEs and RBs are indexed by $u \in \mathcal{U} \triangleq \{1, \dots, U\}$ and $r \in \mathcal{R} \triangleq \{1, \dots, R\}$, respectively, with $R<U$. The HAPS is equipped with a uniform planar array (UPA), facing towards ground, consisting of $N_{\text{x}} \times N_{\text{y}}$ antenna elements, where $N_\text{x}$ and $N_\text{y}$ denote the number of antenna elements along the $x$ and $y$-axes, respectively, as depicted in Fig.~\ref{systemmodel}. The system operates in the sub-6 GHz frequency band and assumes equal bandwidth allocation of $B$~Hz to each UE.

Due to the high altitude of HAPS, the HAPS--UE links are typically dominated by a strong LoS component, which varies slowly over time. Hence, we adopt a large-scale channel model in this work. For UE $u$, let $d_u$ denote the HAPS--UE distance and let $(\theta_u,\phi_u)$ be the elevation and azimuth angles of departure from the HAPS toward UE $u$, where $\theta_u \in [0,\pi/2]$ and $\phi_u \in [-\pi,\pi)$ (as shown in Fig.~\ref{systemmodel}). Using the UPA, HAPS can create $L$ directional beams in ground regions, referred to as clusters. Each cluster is indexed as $\ell \in$ $\mathcal{L}\triangleq \{1,\dots,L\}$, and it covers multiple UEs within its coverage area. Let $(\theta_\ell, \phi_\ell)$ denote the elevation and azimuth angles defining the boresight of the beam associated with cluster $\ell$. We denote the beam gain of cluster $\ell$ toward UE $u$ by $g_{\ell,u}$, defined~as
\begin{align}
g_{\ell,u} = G_{\ell}(\theta_u,\phi_u) \, \text{PL}_u,~\forall \ell \in \mathcal{L},~\forall u \in \mathcal{U},
\end{align}
where $\text{PL}_u=\Big({4\pi d_u f_c}/{c}\Big)^2$ denotes the free-space path loss (FSPL) between HAPS and UE $u$, where $f_c$ is the carrier frequency and $c$ is the speed of light. $G_{\ell}(\theta_u,\phi_u)$ represents the effective antenna gain of cluster $\ell$ toward UE $u$ located at the angular position $(\theta_u,\phi_u)$, obtained from the composite antenna radiation pattern, as specified in the International Telecommunication Union (ITU) recommendation report \cite[Tables~3 \& 4]{itur_m2101_2017}. 
We define $\mathcal{U}_\ell$ as the set of users served by cluster $\ell$. Each UE $u$ is associated with exactly one cluster, i.e.,
$\mathcal{U}_\ell \cap \mathcal{U}_{\ell'} = \emptyset,\ \forall \ell \neq \ell'$.
Moreover, each UE $u$ is allocated only one RB, denoted as $r_u$. Since the number of RBs is limited ($R < U$), multiple UEs inevitably share the same RB, leading to intra-RB interference. To address this challenge, in the following section we propose an efficient algorithm that jointly performs UE clustering and RB allocation, with the objective of minimizing interference.

\section{Angular-Aware Clustering and Interference-Aware RB Allocation}\label{Sec:Clustering and RB Allocation}
In this section, we propose an algorithm that clusters UEs based on their angular similarity, steers one beam toward each cluster $\ell$ to serve its associated UEs, and allocates the $R$ available orthogonal RBs among the $U$ UEs. Based on the antenna pattern in \cite{itur_m2101_2017}, the effective antenna gain from cluster $\ell$ toward UE $u$, denoted by $g_{\ell,u}$, is a deterministic function of the UE's angular position $(\theta_u,\phi_u)$. Consequently, UEs with proximate angular positions experience comparable antenna gains. Therefore, scheduling such UEs on the same RB results in severe inter-UE interference. To mitigate this effect, we adopt an angular-aware clustering and interference-aware RB assignment strategy that relies only on UE angular position. 

In this strategy, we set the number of clusters to $L \triangleq \left\lceil U/R \right\rceil$, where $\lceil \cdot \rceil$ is a ceiling operator. This approach limits the maximum number of UEs per cluster $\leq R$, and ensures orthogonal RBs allocation within each cluster. Next, we let each UE $u$ to be represented by the feature vector $\mathbf{x}_u = [\theta_u, \cos\phi_u, \sin\phi_u]^\top$, which captures its elevation and azimuth directions (the $\cos/\sin$ mapping avoids angle discontinuity at $\pm\pi$). 
Given $\{\mathbf{x}_u\}$, the capacity-constrained $K$-means approach, as presented in~\cite{ConstrainedKMeans}, is used to group the UEs into $L$ clusters while enforcing a maximum cluster size of $R$. As a result, UEs with proximate angular positions are grouped into the same cluster and receive orthogonal RBs allocation, thereby avoiding intra-cluster co-channel interference.
After clustering, the next step is to determine the beam boresight direction for each cluster. To promote intra-cluster fairness, each beam is steered toward the direction that maximizes the minimum effective antenna gain among the UEs in the corresponding cluster $\mathcal{U}_\ell$, i.e., adopting a worst-UE-based beam steering (WU-Clustering) strategy, as detailed in Algorithm~\ref{alg:clustering_beam}.
\begin{algorithm}[h]
\small
\caption{\small Angular-aware clustering with worst-UE beam \mbox{steering} (WU-Clustering)}
\label{alg:clustering_beam}
\begin{algorithmic}[1]
\small
\State \textbf{Input:} $U,~R$, $(\theta_u,\phi_u),~\forall u \in \mathcal{U},$
\State \textbf{Output:} Clusters $\{\mathcal{U}_\ell\}$ and beam directions $(\theta_\ell,\phi_\ell),~\forall \ell \in \mathcal{L}$.

\State Compute $(\theta_u,\phi_u)$ and set $\mathbf{x}_u=[\theta_u,\cos\phi_u,\sin\phi_u]^\top,~\forall u\in\mathcal{U}$.

\State Apply capacity-constrained $K$-means with $K=\lceil U / R \rceil$ and maximum number of UEs per cluster $R$ on $\{\mathbf{x}_u\}$ to obtain $\{\mathcal{U}_\ell\}$

\State $(\theta_\ell,\phi_\ell) \leftarrow \operatorname*{argmax}_{(\theta,\phi)} 
\min_{u \in \mathcal{U}_\ell} g_{\ell,u},~\forall \ell \in \mathcal{L}$.
\end{algorithmic}
\end{algorithm}

After obtaining the UE cluster sets $\mathcal{U}_\ell,~\forall \ell \in \mathcal{L}$, the HAPS forms one beam per cluster $\ell$ and steers it toward the corresponding direction $(\theta_\ell,\phi_\ell)$. Subsequently, given the resulting antenna gains $g_{\ell,u},~\forall \ell,~\forall u$, RBs are allocated to UEs while enforcing intra-cluster orthogonality.
To this end, for each UE $u$, the HAPS first determines the set of candidate RBs $\mathcal{R}'_u$ as the set of RBs not yet used within the serving cluster of $u$. Accordingly, UE $u$ is assigned to the RB that minimizes the accumulated interference from already scheduled UEs, i.e.,
\begin{align}
r_u = \arg\min_{r \in \mathcal{R}'_u} \sum_{j \in \mathcal{U}_r} g_{\ell(j),u}, 
\quad \forall u \in \mathcal{U},
\end{align}
where $\mathcal{U}_r$ denotes the set of UEs co-assigned to RB $r$, and $\ell(j)$ denotes the index of the cluster serving UE $j$.

In the proposed angular-aware WU-clustering and interference-aware RB allocation, $U$ UEs are first grouped into $L$ clusters, and then RBs are assigned to minimize the aggregated leakage toward each newly scheduled UE. Since RBs are reused in different clusters, intra-RB interference is inevitable. To manage this inter-cluster interference, we apply RSMA on a per-RB basis, as described in the following section.

\section{HAPS Inter-Cluster Interference Management through RSMA}\label{Sec:RSMA}
In this section, we propose an RSMA-based transmission scheme to manage interference within each RB caused by non-orthogonal RB allocation among UEs from different clusters. To this end, let $W_u$ denote the message intended for UE $u$, which is split into a common part $W_u^{(c)}$ and a private part $W_u^{(p)}$. For each RB $r$, the HAPS combines the common parts of UEs co-scheduled on RB $r$, i.e., $\{W_u^{(c)}: u \in \mathcal{U}_r\}$, into a single RB-specific common message $W_r^{(c)}$, which is encoded into a common stream $s_r^{(c)}$. Accordingly, the private part $W_u^{(p)}$ is independently encoded into a private stream $s_u^{(p)}$ for each UE $u \in \mathcal{U}$. The common stream $s_r^{(c)}$ is transmitted with power $p_{r,\ell}^{(c)},~\forall r,~\forall \ell$. 
In addition, each private stream $s_u^{(p)}$ is transmitted on its corresponding cluster beam $\ell(u)$ with power $p_u^{(p)}$ over its assigned RB $r_u$. We assume unit power for all transmitted streams, i.e., $\mathbb{E}[|s_r^{(c)}|^2] = 1,~\forall r$ and $\mathbb{E}[|s_u^{(p)}|^2] = 1,~\forall u$. Accordingly, the received signal by UE $u$, denoted as $y_u$, can be written as
\begin{equation}\label{eq:rsma_rx}
 y_{u} = \underbrace{\sum_{\ell\in \mathcal{L}}\sqrt{p_{r_u,\ell}^{(c)} g_{\ell,u}}\, s_{r_u}^{(c)}}_{\text{received common part}}
 + \underbrace{\sum_{\substack{k\in\mathcal{U}_{r_u}}}\sqrt{p_{k}^{(p)} g_{\ell(k),u}}\, s_{k}^{(p)}}_{\text{received private part}}
 + n_{u},~\forall u\in\mathcal{U},
\end{equation}
where $n_{u}\sim\mathcal{CN}(0,\sigma_n^2)$ denotes additive white Gaussian noise (AWGN) with variance $\sigma_n^2$.
After receiving the signal $y_u$, UE $u$ first decodes the common stream $s_{r_u}^{(c)}$ by treating all private streams transmitted on RB $r_u$ as noise. Accordingly, the corresponding SINR of the common stream at UE $u$ can be expressed as
\begin{equation}\label{eq:sinr_common}
\gamma^{(c)}_{u}=
\frac{\sum\limits_{\ell\in \mathcal{L}} p_{r_u,\ell}^{(c)} g_{\ell,u}}
{\sum\limits_{\substack{k\in\mathcal{U}_{r_u}}} p_{k}^{(p)}\, g_{\ell(k),u} + \sigma_n^2},~\forall u\in\mathcal{U},
\end{equation}
and the achievable common data rate at UE $u$ as
\begin{equation}\label{eq:rate_common_user}
R^{(c)}_{u}=B \, \log_2\big(1+\gamma^{(c)}_{u}\big),~\forall u\in\mathcal{U}.
\end{equation}

Since $s_{r}^{(c)}$ must be decoded by all UEs co-scheduled on RB $r$, the common stream rate on RB $r$ is limited by
\begin{equation}\label{eq:rate_common}
R^{(c)}_{r}=\min\limits_{u\in\mathcal{U}_r} R^{(c)}_{u},~\forall r\in\mathcal{R}.
\end{equation}

Let $C_u \ge 0$ denote the portion of the common rate $R^{(c)}_{u}$ allocated to UE $u$. Then, for each RB $r$, we will have:
\begin{equation}\label{eq:common_alloc}
\sum\limits_{u\in\mathcal{U}_r} C_{u} \le R^{(c)}_{r},~\forall r\in\mathcal{R}.
\end{equation}

After decoding $s_{r_u}^{(c)}$, UE $u$ performs successive interference cancellation (SIC) to subtract the common stream.
After SIC, UE $u$ decodes its private stream $s_{u}^{(p)}$ while treating the other private streams on RB $r_u$ as noise. As a result, the corresponding SINR will be as
\begin{equation}\label{eq:sinr_private}
\gamma^{(p)}_{u}=
\frac{p^{(p)}_{u} g_{\ell(u),u}}
{\sum\limits_{\substack{k\in\mathcal{U}_{r_u}\\k\neq u}} p_{k}^{(p)}\, g_{\ell(k),u}+\sigma_n^2},~\forall u\in\mathcal{U},
\end{equation}
and the achievable private data rate at UE $u$ will be as
\begin{equation}\label{eq:rate_private}
R^{(p)}_{u}=B \, \log_2\big(1+\gamma^{(p)}_{u}\big),~\forall u\in\mathcal{U}.
\end{equation}

The proposed RSMA scheme provides additional degrees-of-freedom via per-cluster power allocation to each RB, i.e., $p_{r,\ell}^{(c)},~\forall r \in \mathcal{R},~\forall \ell \in \mathcal{L},$ to manage the inter-cluster interference. However, for RSMA to perform efficiently, proper power allocation between the common and private streams is required, along with the allocation of the common rate for each UE, denoted by $C_u,~\forall u \in \mathcal{U}$. As a result, $p_{r,\ell}^{(c)},~\forall r \in \mathcal{R},~\forall \ell \in \mathcal{L},$ and $p^{(p)}_{u},~\forall u \in \mathcal{U},$ should be designed under the total available transmit power constraint at the HAPS, given by
\begin{equation}\label{eq:power_rsma}
\sum\limits_{r\in \mathcal{R}}\sum\limits_{\ell\in\mathcal{L}} p_{r,\ell}^{(c)} 
+ \sum\limits_{u\in\mathcal{U}} p_{u}^{(p)} 
\le P_{\mathrm{T}},
\end{equation}
where $P_{\mathrm{T}}$ denotes the total available transmit power budget at the HAPS. To this end, we formulate an optimization problem to jointly design the common and private power allocation, as well as the common rate assigned to each UE, with the objective of maximizing the minimum achievable UE data rate.

\section{Max--Min Fair RSMA Power Allocation (MMF-RSMA-PA) Design}\label{Sec:MMFProblem}
\subsection{Problem Formulation}
Using the achievable common and private rates of UE $u$, defined in \eqref{eq:rate_common_user} and \eqref{eq:rate_private}, respectively, the total achievable data rate of UE $u$ is given by
\begin{equation}\label{eq:user_total_rate}
R_{u} = C_u + R_{u}^{(p)},\ \forall u\in\mathcal{U}.
\end{equation}

Accordingly, the MMF RSMA power allocation problem can be formulated as follows:
\begin{subequations}\label{eq:mmf_problem}
\begin{align}
\small
\max_{\mathbf{P}^{(c)},\,\mathbf{p}^{(p)},\,\mathbf{C}} \;& \min_{u\in\mathcal{U}} R_u \label{Objective_MMF}\\
\text{s.t.}\quad
& \eqref{eq:common_alloc},~\eqref{eq:power_rsma}, \\
& \mathbf{C} \geq 0,~\mathbf{p}^{(p)} \geq 0,~\mathbf{P}^{(c)} \geq 0, \label{Const:positive_Var}
\end{align}
\end{subequations}
where matrix $\mathbf{P}^{(c)}$ and vectors $\mathbf{p}^{(p)}$ and $\mathbf{C}$ collect the variables $p_{r,\ell}^{(c)},~\forall r \in \mathcal{R},~\forall \ell \in \mathcal{L},$ $p^{(p)}_{u},~\forall u \in \mathcal{U},$ and $C_u,~\forall u \in \mathcal{U},$ respectively.
In problem~\eqref{eq:mmf_problem}, constraint~\eqref{eq:common_alloc} ensures that the common stream can be successfully decoded by all UEs co-scheduled on the same RB, while constraint~\eqref{eq:power_rsma} ensures that the total transmit power allocated to common and private streams does not exceed the available HAPS power budget $P_{\mathrm{T}}$. Problem~\eqref{eq:mmf_problem} is non-convex due to the objective function in \eqref{Objective_MMF} and constraint~\eqref{eq:common_alloc}, which depend on the SINR expressions in \eqref{eq:sinr_common} and \eqref{eq:sinr_private}, involving ratios of optimization variables. Therefore, in the remainder of this section, we adopt a successive convex approximation (SCA) approach to transform problem~\eqref{eq:mmf_problem} into a tractable and convex form that can be solved iteratively to obtain a suboptimal solution.
\vspace{-3mm}
\subsection{Proposed Solution}
First, we address the objective function \eqref{Objective_MMF}. To this end, we introduce an epigraph variable $t$ to replace the objective term $\min_{u\in\mathcal{U}} R_u$, where $t$ represents the minimum achievable UE data rate. Accordingly, we impose the constraint
\begin{align}\label{eq:MMF_Constraint}
R_u \ge t,\ \forall u \in \mathcal{U}.
\end{align}

To handle the non-convexity of this constraint, we introduce slack variables $\xi_u^{(p)},\ \forall u \in \mathcal{U}$, as lower bounds for the private SINR $\gamma_u^{(p)}$. Then, the constraint can be rewritten as
\begin{align}\label{eq:MMF_Constraint_V2}
C_u + B \, \log_2\big(1 + \xi_u^{(p)}\big) \ge t,\ \forall u \in \mathcal{U},
\end{align}
which represents a convex constraint. Now, to ensure that $\xi_u^{(p)} \le \gamma_u^{(p)}$ holds for all $u$, we introduce an auxiliary slack variable $\beta_u^{(p)}$ and introduce the following constraints:
\begin{IEEEeqnarray}{lCl}
\xi_u^{(p)} \, \beta_u^{(p)} \le p_u^{(p)} \, g_{\ell(u),u},\ \forall u \in \mathcal{U}, \qquad \IEEEyesnumber \IEEEyessubnumber* \label{eq:xibeta_private}\\
\beta_u^{(p)} \ge \sum_{\substack{k\in\mathcal{U}_{r_u}\\ k \neq u}} p_k^{(p)} \, g_{\ell(k),u} + \sigma_n^2,\ \forall u \in \mathcal{U}. \label{eq:beta_private}
\end{IEEEeqnarray}

Constraint \eqref{eq:beta_private} represents a convex constraint but constraint \eqref{eq:xibeta_private} is nonconvex due to the product of two variables in the left-hand side. To handle this, we move $\beta^{(p)}_u$ to the right-hand side (RHS) and replace the RHS with its first-order Taylor approximation as
\begin{equation}\label{eq:taylor_private}
\xi_u^{(p)} \le 
\frac{p_u^{(p,n)} g_{\ell(u),u}}{\beta_u^{(p,n)}}
+ \frac{g_{\ell(u),u}}{\beta_u^{(p,n)}}\left(p_u^{(p)} - p_u^{(p,n)}\right)
- \frac{p_u^{(p,n)} g_{\ell(u),u}}{\left(\beta_u^{(p,n)}\right)^2}
\left(\beta_u^{(p)} - \beta_u^{(p,n)}\right),
\ \forall u \in \mathcal{U},
\end{equation}
where $p_u^{(p,n)}$ and $\beta_u^{(p,n)}$ denote the values of variables $p_u^{(p)}$ and $\beta_u^{(p)}$ at SCA iteration $n$. 
Next, we deal with the constraint \eqref{eq:common_alloc}. To this end, considering \eqref{eq:rate_common_user}, first we replace the right-hand side with $R^{(c)}_u$ as
\begin{equation}
    \sum\limits_{u\in\mathcal{U}_r} C_{u} \le R^{(c)}_{j},~\forall~r\in\mathcal{R},~\forall~j\in\mathcal{U}_r.
\end{equation}

Accordingly, to tackle the nonconvexity of the RHS, similar to the private part, we define slack variables $\xi^{(c)}_u$ and $\beta_u^{(c)}$ as
\begin{IEEEeqnarray}{lCl}
 \xi^{(c)}_u \beta_u^{(c)}  &\le \sum_{\ell\in \mathcal{L}} p_{r_u,\ell}^{(c)} g_{\ell,u},~\forall u \in \mathcal{U},\qquad \IEEEyesnumber \IEEEyessubnumber* \label{eq:xibeta_common}\\
 \beta_u^{(c)} &\ge \sum\limits_{\substack{k\in\mathcal{U}_{r_u}}} p_{k}^{(p)}\, g_{\ell(k),u} + \sigma_n^2,~\forall u \in \mathcal{U},
\label{eq:beta_common}
\end{IEEEeqnarray}
which similar to the private part, the constraint \eqref{eq:xibeta_common} can be converted to approximated convex form as
\begin{equation}\label{eq:taylor_common}
\xi_u^{(c)} \le 
\frac{\sum\limits_{\ell\in \mathcal{L}} p_{r_u,\ell}^{(c,n)} g_{\ell,u}}{\beta_u^{(c,n)}}
+ \sum\limits_{\ell\in \mathcal{L}} \frac{g_{\ell,u}}{\beta_u^{(c,n)}} \left(p_{r_u,\ell}^{(c)} - p_{r_u,\ell}^{(c,n)}\right)
- \frac{\sum\limits_{\ell\in \mathcal{L}} p_{r_u,\ell}^{(c,n)} g_{\ell,u}}{\left(\beta_u^{(c,n)}\right)^2}
\left(\beta_u^{(c)} - \beta_u^{(c,n)}\right),
\ \forall u \in \mathcal{U}.
\end{equation}

Accordingly, constraint \eqref{eq:common_alloc} will be transformed to
\begin{equation}\label{eq:CommonAllocation}
    \sum\limits_{u\in\mathcal{U}_r} C_{u} \le B \, \log_2(1+\xi^{(c)}_j),~\forall~r\in\mathcal{R},~\forall~j\in\mathcal{U}_r,
\end{equation}
and the approximated optimization problem at SCA iteration $n$ will be as
\begin{subequations}\label{eq:mmf_problem_approx}
\begin{align}
\small
\max_{\substack{\mathbf{P}^{(c)},\mathbf{p}^{(p)},\mathbf{C},\\ \boldsymbol{\xi}^{(c)},\boldsymbol{\xi}^{(p)},\boldsymbol{\beta}^{(c)},\boldsymbol{\beta}^{(p)},t}} \;& ~t \label{Objective_MMF_V2} \\
\text{s.t.}\quad
& \eqref{eq:power_rsma},~\eqref{Const:positive_Var},~\eqref{eq:MMF_Constraint_V2}, \nonumber\\
& \eqref{eq:beta_private},~\eqref{eq:taylor_private},~\eqref{eq:beta_common},~\eqref{eq:taylor_common},~\eqref{eq:CommonAllocation}. \nonumber
\end{align}
\end{subequations}

At each SCA iteration $n$, we solve the resulting convex problem~\eqref{eq:mmf_problem_approx} (e.g., using CVX), update the linearization point, and stop when the relative improvement in $t$ is below a preset threshold. The proposed iterative MMF-RSMA-PA design algorithm is described below.
\begin{algorithm}[h!]
\small
\caption{\small Iterative MMF-RSMA-PA design algorithm}
\label{alg:alg1}
\begin{algorithmic}[1]
\small
\State \textbf{Input:} $U,~R,~\mathcal{U}_\ell,~\forall \ell,~\mathcal{U}_r,~\forall r,~P_T,~g_{\ell,u},~\forall \ell,~\forall u,~\sigma_n^2,~N_{\text{iter}}$.
\State \textbf{Output:} ${\mathbf{P}^{(c)}}^*,~{\mathbf{p}^{(p)}}^*,~\mathbf{C}^*$.

\State Initialize $\boldsymbol{\beta}^{(c,0)},~\boldsymbol{\beta}^{(p,0)},~\mathbf{P}^{(c,0)},~\mathbf{p}^{(p,0)}$ and set $n \leftarrow 0$.

\Repeat
    \State Solve (\ref{eq:mmf_problem_approx}) to obtain $\mathbf{P}^{(c,n)*},~\mathbf{p}^{(p,n)*},~\boldsymbol{\beta}^{(n)*},~t^{(n)}$.
    
    \State Update:
    \State \hspace{1em} $\boldsymbol{\beta}^{(c,n+1)} \leftarrow \boldsymbol{\beta}^{(c,n)*}, \quad
                         \boldsymbol{\beta}^{(p,n+1)} \leftarrow \boldsymbol{\beta}^{(p,n)*}$,
    \State \hspace{1em} $\mathbf{P}^{(c,n+1)} \leftarrow \mathbf{P}^{(c,n)*}, \quad
                         \mathbf{p}^{(p,n+1)} \leftarrow \mathbf{p}^{(p,n)*}$,
    \State \hspace{1em} $n \leftarrow n + 1$.
\Until{$\left|t^{(n)} - t^{(n-1)}\right| / t^{(n-1)} \leq \epsilon $ \textbf{or} $n \geq N_{\text{iter}}$}.
\end{algorithmic}
\end{algorithm}
\section{Numerical Results}\label{Sec:SimResults}
We consider a single HAPS located at an altitude of $20$~km serving $U=60$ single-antenna UEs over $R=10$ RBs. The UEs are uniformly distributed over a $2$~km radius circular geographic area. The HAPS is equipped with a UPA antenna with configuration of $N_{\text{x}} \times N_{\text{y}} = 8 \times 8$, used to form one beam per cluster of UEs. The carrier frequency is $f_c = 2.545$~GHz. The remaining simulation parameters and their values are summarized in Table~\ref{tab:SimulationParameters}.
To evaluate the performance of the proposed scheme, we consider the following three scenarios:
\begin{itemize}
    \item \textit{Scenario~1 (WU-Clustering + RSMA):} This scenario corresponds to the proposed algorithm, where clustering is performed using the WU-clustering Algorithm~\ref{alg:clustering_beam}, and power allocation design is obtained via MMF-RSMA-PA Algorithm~\ref{alg:alg1}.
    \item \textit{Scenario~2 (C-Clustering + RSMA):} In this scenario, UEs are clustered based on their angular similarity but beams are steered toward the centroid of each cluster $\ell$, instead of maximizing the worst-UE gain. The power allocation design is performed using MMF-RSMA-PA Algorithm~\ref{alg:alg1}.
    \item \textit{Scenario~3 (WU-Clustering without RSMA):} This scenario follows the same clustering mechanism as Scenario~1, but without RSMA, i.e., only private streams are transmitted.
\end{itemize}
\begin{table}[!t]
\caption{\small Simulation Parameters.}\label{tab:SimulationParameters}
\centering
\begin{tabular}{|c||c|}
\hline
\textbf{Parameter} & \textbf{Value}\\
\hline
Antenna elements spacing,~$d_x,~d_y$ & $\lambda/2$\\
\hline
AWGN noise variance,~$\sigma^2_n$ & $-100$ dBm\\
\hline
HAPS total transmit power,~$P_{T}$ & $55$ dBm~\cite{itu_wrc}\\
\hline
HAPS $3$\,dB beamwidth & $65^{\circ}$ \cite{itu_wrc}\\
\hline
HAPS antenna element gain & $8$ dBi\cite{itu_wrc}\\
\hline
HAPS antenna element front to back ratio & $30$ dB\cite{itu_wrc}\\
\hline
Maximum number of SCA iterations ($N_{\text{iter}}$) & $20$\\
\hline
\end{tabular}
\end{table}
\begin{figure}
    \centering
    \includegraphics[width=0.6\linewidth]{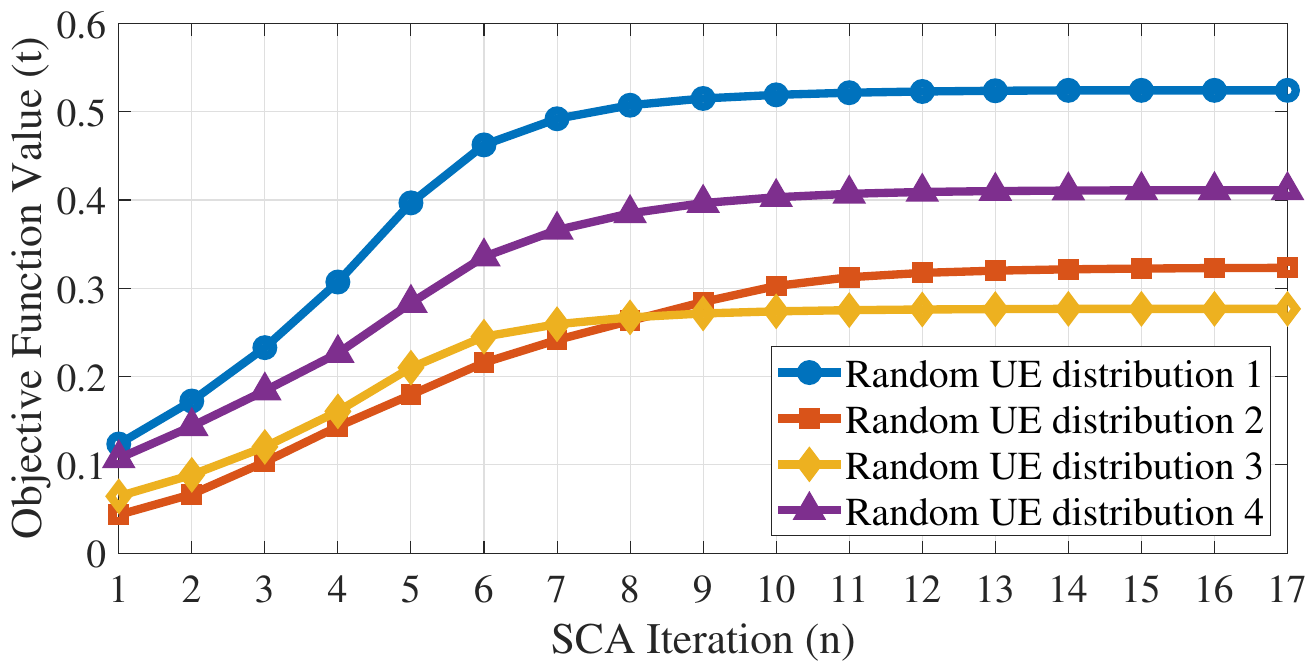}
    \caption{\small Convergence behavior of Algorithm~\ref{alg:alg1}.}
    \label{fig:Convergence}
\end{figure}

The RB allocation in all scenarios follows the interference-aware approach, as described in Section~\ref{Sec:Clustering and RB Allocation}. All results are averaged over $1000$ independent network realizations. The algorithm is implemented using CVX parser, which runs on MATLAB, with \texttt{MOSEK} 9.1.9 as an internal solver.

First, we evaluate the convergence behavior of the proposed Algorithm~\ref{alg:alg1} in Fig.~\ref{fig:Convergence}, which shows the evolution of the objective function value $t$ over SCA iterations $n$ for four independent random UE distributions. The results demonstrate a monotonic increase in the MMF objective $t$, which quickly saturates to a stable value within fewer than $13$ iterations across all realizations. While different UE distributions lead to varying initial values and convergence paths, the objective function values converge quickly in all four distributions.

\begin{figure}[t]
    \centering
    \includegraphics[width=0.6\linewidth]{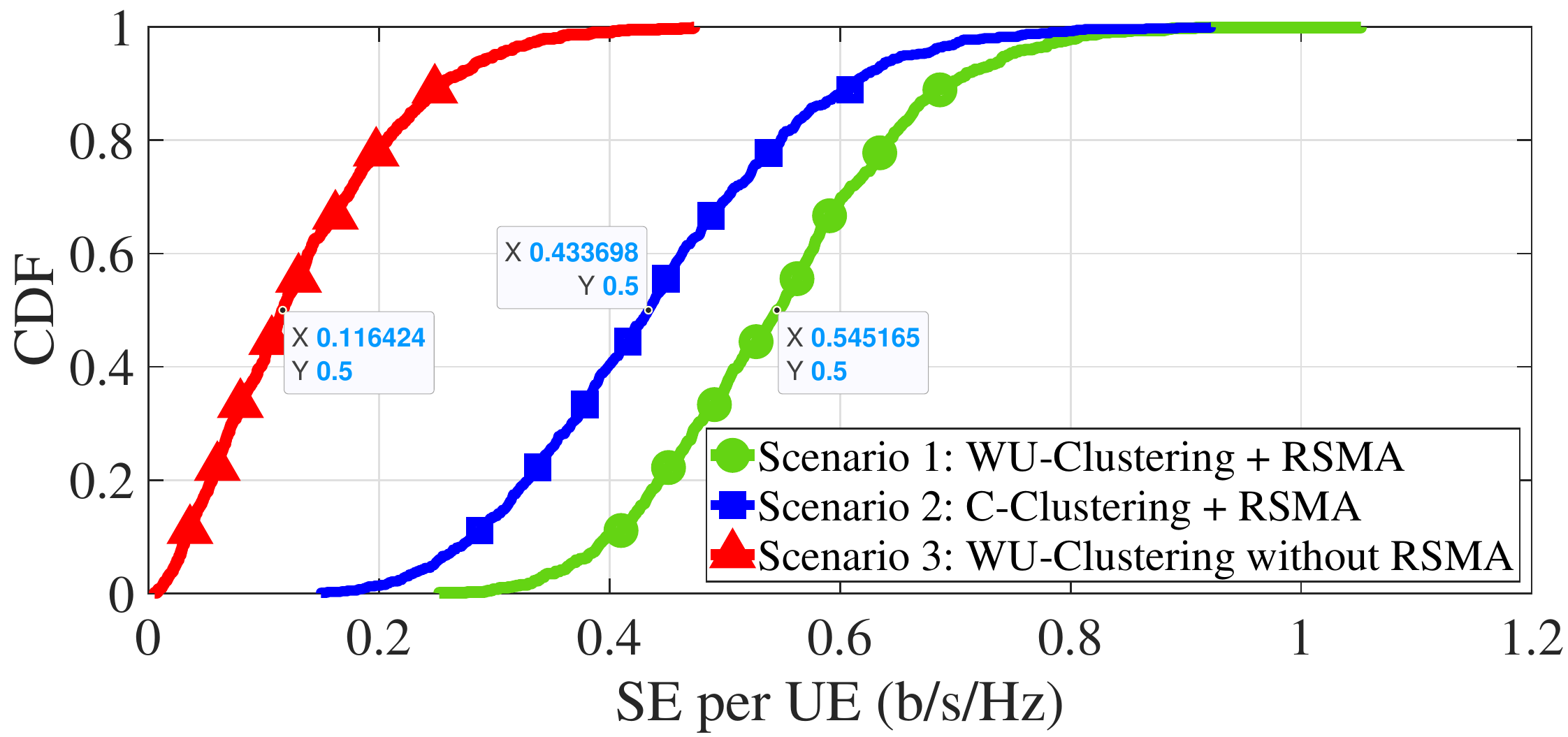}
    \caption{\small CDF of SE per-UE (b/s/Hz) for the three scenarios.}
    \label{fig:CDF_SE}
\end{figure}

Fig.~\ref{fig:CDF_SE} presents and compares the cumulative distribution function (CDF) of the SE per UE (i.e., $R_u/B,~\forall u$) for the three scenarios described above. Comparing Scenario~1 (WU-Clustering + RSMA) with Scenario~3 (WU-Clustering without RSMA) reveals that incorporating RSMA for intra-RB interference management yields a substantial performance gain. In particular, the median SE (i.e., the 50th percentile) improves from approximately $0.12$~b/s/Hz to $0.55$~b/s/Hz, corresponding to more than a fourfold increase. This highlights the effectiveness of RSMA in mitigating inter-cluster interference caused by RB reuse.
Further, comparing Scenarios 1 and 2, it can be observed that the proposed WU-clustering scheme consistently outperforms the centroid-based approach. This gain stems from the fact that the WU-clustering algorithm explicitly accounts for the worst-UE channel conditions when designing beam directions, thereby improving fairness and enhancing the minimum achievable rates across UEs. In contrast, the centroid-based strategy does not adapt to UE performance, leading to degraded SE, particularly for UEs in unfavorable channel conditions.
Overall, the proposed WU-clustering + RSMA scheme significantly improves the SE distribution across all percentiles, demonstrating superior performance compared to the considered benchmark schemes.

\begin{figure}[t]
    \centering
    \includegraphics[width=0.6\linewidth]{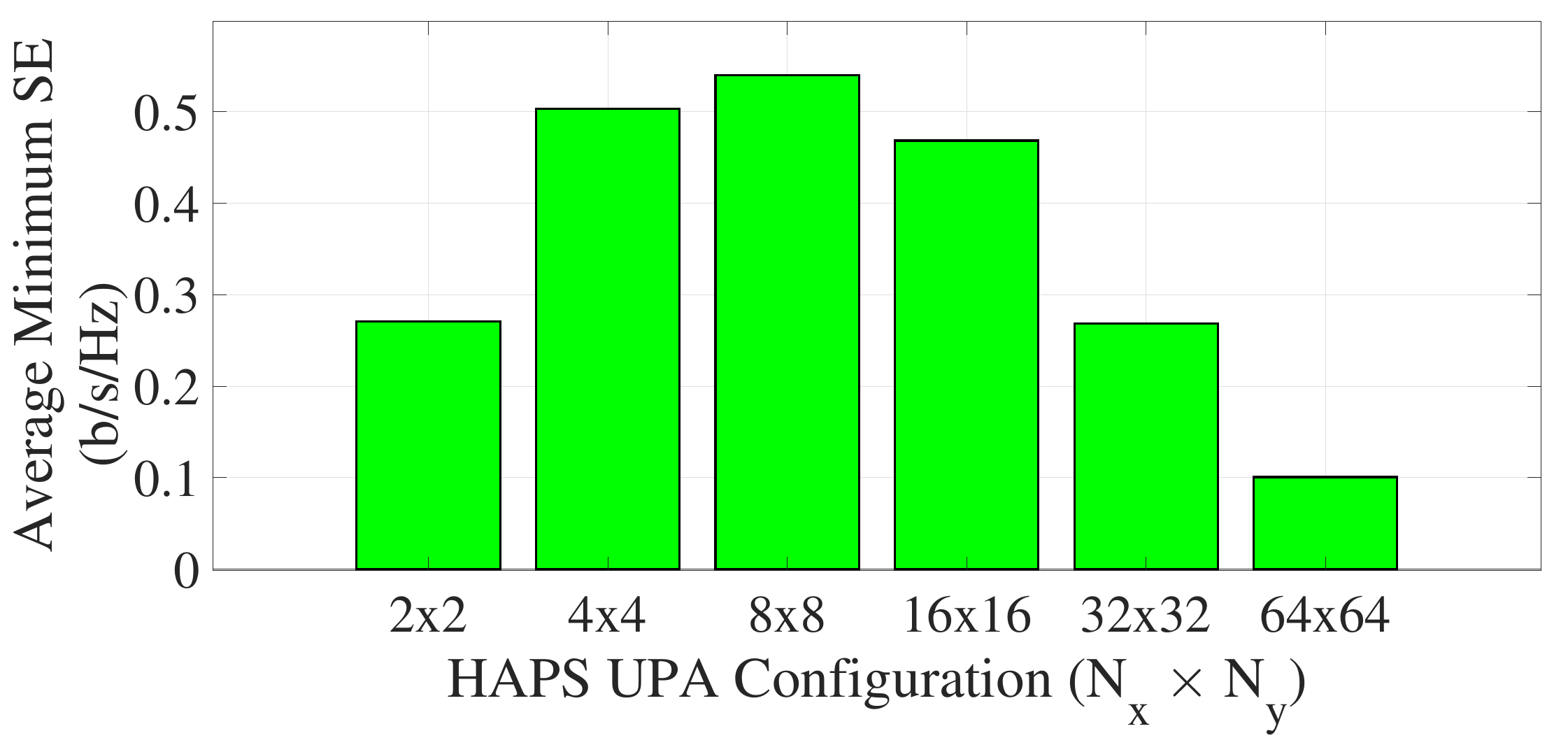}
    \caption{\small Impact of HAPS UPA configuration on the average minimum SE (b/s/Hz) in the proposed algorithm.}
    \label{fig:HAPSAntenna}
\end{figure}
Fig.~\ref{fig:HAPSAntenna} depicts the average minimum SE for different HAPS UPA configurations. As the number of antenna elements increases, the formed beams become narrower. While narrower beams enhance the beamforming gain toward UEs located near the beam center, they reduce the gain for UEs located farther away from the main lobe. 
At the same time, narrower beams limit the leakage of interference toward other clusters, thereby improving inter-cluster interference management. This leads to a fundamental trade-off between beamforming gain and interference mitigation.
As observed, increasing the antenna size up to $8 \times 8$ improves performance due to more effective interference suppression. However, further increasing the number of antenna elements results in performance degradation, as the reduced coverage of narrower beams negatively impacts UEs located away from the beam center. 
\section{Conclusion} \label{Sec:Conclusion}
This paper investigated interference management in HAPS networks for a multi-beam downlink system with limited orthogonal RBs. By leveraging angular-aware UE clustering and worst-UE beam steering, we proposed an efficient framework to manage the interference. To further mitigate inter-cluster interference arising from RB reuse, an RSMA-based transmission strategy was incorporated, enabling flexible power and rate allocation across common and private streams. Simulation results demonstrated that the proposed approach significantly improves per-UE spectral efficiency, confirming its effectiveness.
\bibliographystyle{IEEEtran}
\bibliography{references}

\end{document}